\documentclass[twocolumn,showpacs,preprintnumbers,amsmath,amssymb]{revtex4}
\usepackage{amsfonts}
\usepackage{amsmath}
\usepackage{graphicx}
\usepackage{dcolumn}
\usepackage{bm}
\usepackage{overpic}
\usepackage{booktabs}
\usepackage{color}
\def\degree{\hbox{$^\circ$}}

\begin{document}
\title{The dominant imprint of Rossby waves in the climate network}
\author{Yang Wang$^1$\footnote{wangyang.maple@gmail.com}, Avi Gozolchiani$^2$, Yosef Ashkenazy$^3$, Yehiel Berezin$^1$, Oded Guez$^1$, and Shlomo Havlin$^1$\footnote{havlin@ophir.ph.biu.ac.il}}

\affiliation{1 Department of Physics, Bar-Ilan University, Ramat-Gan
  52900, Israel\\
  2 Institute of Earth Sciences, The Hebrew University of Jerusalem,
  Jerusalem, 91904, Israel\\
  3 Department of Solar Energy and Environmental Physics, BIDR,
  Ben-Gurion University, Midreshet Ben-Gurion, 84990, Israel }

\date{\today}
\begin{abstract}
  The connectivity pattern of networks, which are based on a correlation
  between ground level temperature time series, shows a dominant dense
  stripe of links in the southern ocean. We show that statistical
  categorization of these links yields a clear association with the
  pattern of an atmospheric Rossby wave, one of the major mechanisms
  associated with the weather system and with planetary scale energy
  transport. It is shown that alternating densities of negative and
  positive links (correlations) are arranged in half Rossby wave
  distances around 3,500 km, 7,000 km and 10,000 km and are aligned
  with the expected direction of energy flow, distribution of time
  delays and the seasonality of these waves. It is also shown that
  long distance links (i.e., of distances larger than 2,000 km) that
  are associated with Rossby waves are the most dominant in the
  climate network. Climate networks may thus be used as an efficient
  new way to detect and analyze Rossby waves, based on reliable and
  available ground level measurements, in addition to the frequently
  used 300 hPa reanalysis meridional wind data.
\end{abstract}

\maketitle

Networks have become an important tool for analyzing technological and
natural
systems~\cite{Boccaletti2006,Newman2010,Cohen2010,Albert2002}. Examples
range from social relations~\cite{Newman2002}, biochemical
interactions~\cite{Guimera2005,Palla2005,Jeong2000}, information flow
through the world wide web~\cite{Huberman2001}, physiological
activities~\cite{Bashan2012}, and the mitigation of attacks on
transportation infrastructures~\cite{Schneider2011}. It was suggested
in past years that climate variables, like temperature and
geopotential height, can be viewed as a climate
network~\cite{Tsonis2006,Yamasaki2008,Tsonis2008,Donges2009(1),Berezin2012,Gozolchiani2011,Gurz2012,Donges2009}. In
this representation, different regions of the world are regarded as
nodes of the network, and links of the network represent
communications between different locations via, e.g., heat and
material exchange. A multitude of statistical analysis methods are
often used to capture major variability patterns in climate time
series~\cite{Storch2003,Robinson2008,Alpert2011}, where the correlation matrix plays an important part. The climate network provides a complementary tool to study the
statistical properties of the climate system.

The climate network often has very strong links which are caused by a
proximity (distance) effect~\cite{Berezin2012}. Namely, pairs of sites
close to each other below some threshold distance (typically 2,000 km)
are often strongly positively correlated. A significant fraction of
the stable network structure may be associated with the proximity
effect. It is hence common to analyze climate time series also based
on negative correlations
(e.g.,~\cite{Wallace1981,Walker1925,Hoskins1981}), which usually
represent more interesting remote interactions, called teleconnections.

However, recent studies have not distinguished between positive and
negative correlations in climate
networks~\cite{Bashan2012,Tsonis2006,Yamasaki2008,Tsonis2008,Donges2009(1),Berezin2012,Donges2009,Gurz2012,Gozolchiani2011}. In
particular, it is apparent, based on previous studies, that a large
fraction of the links in the climate network
resides in the southern
ocean~\cite{Yamasaki2008,Gozolchiani2011,Tsonis2008}. This fraction
may include (beyond the proximity effect distance) both negative and
positive correlations.

Here we analyze separately the negative and positive correlations of
the climate network. We show that these links alternate, as a function
of distance, between negative and positive, consistent with a wave
pattern. We find that the time delay associated with these links
increases from one to five days as a function of the distance between
the nodes (the length of the links).  We also analyze the typical
length scale of the links, their seasonality, and the geographical
structure of the climate network; all of these are found here to be
consistent with atmospheric Rossby waves, one of the most efficient
climate mechanisms of planetary-scale energy
transfer~\cite{Chang2005}. Studies of atmospheric Rossby waves are
usually based on the 300 hPa meridional wind velocity reanalysis
data~\cite{Chang1993,Chang1999(1),Chang1999}. Here we show that it is
possible to uncover the characteristics of Rossby waves using more
common and reliable surface data, like surface air temperature. We
find that Rossby waves dominate the climate network, an observation
that, surprisingly, has not been previously reported.


Here, we analyze the daily data of air temperature, sea level
pressure, geopotential height, and meridional velocity
fields. Specifically, we analyze a network of 726 nodes around the
globe (see~\cite{Gozolchiani2011} or small dots in Fig. 4) from the
NCEP/NCAR reanalysis I grid~\cite{Kalnay1996,comment1}. Below, we
mainly focus on the surface temperature field as one of the most
common and reliable types of data. For each node (i.e.,
longitude/latitude grid point) of the network, daily values within the
period 1948-2010 are used, from which we extract anomaly
values. Specifically, if we take the record of a given site in the
grid to be ${\tilde T^y}(d)$, where $y$ is the year and $d$ is the day
(from 1 to 365), then the filtered record is denoted by ${T^y}(d) =
{\tilde T^y}(d) - \frac{1}{N}\sum_y {\tilde {T^y}(d)}$, where $N$ is
the number of years available in the record. We also define
${\Theta _s}(d) \equiv {{{\left[ {{T_s}(d) - \left\langle {{T_s}(d)}
          \right\rangle } \right]} \mathord{\left/ {\vphantom {{\left[
                {{T_s}(d) - \left\langle {{T_s}(d)} \right\rangle }
              \right]} {\left\langle {{{\left( {{T_s}(d) -
                          \left\langle {{T_s}(d)} \right\rangle }
                      \right)}^2}} \right\rangle }}} \right.
      \kern-\nulldelimiterspace} {\left\langle {{{\left( {{T_s}(d) -
                  \left\langle {{T_s}(d)} \right\rangle } \right)}^2}}
      \right\rangle }}^{{1 \mathord{\left/ {\vphantom {1 2}} \right.
        \kern-\nulldelimiterspace} 2}}}$,
where $\langle \cdots \rangle$ is the average of the time series.

\begin{figure}
\center\includegraphics[width=0.5\textwidth]{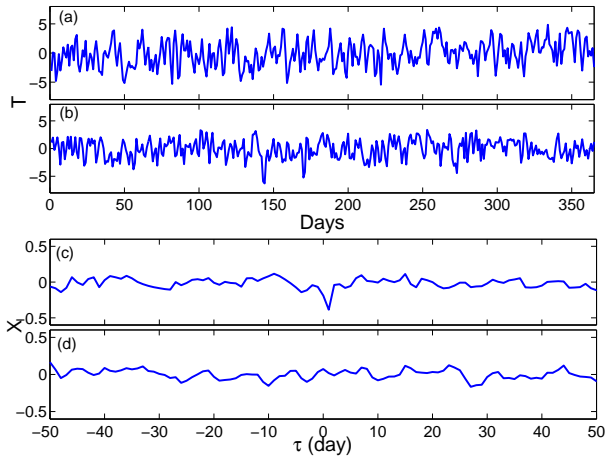}
\caption{(Color online) (a) An example of a seasonally detrended
  surface temperature daily time series from 45\degree{S},
  52.5\degree{E} for the year 1948. (b) Same as (a) for 45\degree{S},
  97.5\degree{E}. (c) The cross-correlation function between the time
  series shown in (a) and (b). (d) Same as (c) but when the time
  series of the two sites are of different years (``shuffling''
  method).}
\label{fig1}
\end{figure}

The link between each pair of sites on the grid, $s_1$ and $s_2$, is
calculated as the cross-correlation function $X_{{s_1},{s_2}}^y(\tau
\ge 0) = \left\langle {\Theta _{{s_1}}^y(d)}{\Theta _{{s_2}}^y(d +
    \tau )} \right\rangle $, where $\tau$ is the time lag and
$X_{{s_1},{s_2}}^y(\tau)=X_{{s_2},{s_1}}^y(-\tau)$. We define the time
lag, $\tau^*$, at which $X_{{s_1},{s_2}}^y(\tau)$ is maximal (or
minimal), as the time delay of a pair $s_1$, $s_2$. When $s_1$ is to
the west of $s_2$ and the time lag is positive, the link direction is
to the east. We distinguish between positive and negative link weights
as follows
\begin{equation}
W_{{s_1},{s_2}}^y = \frac{{{\rm MAX}(X_{{s_1},{s_2}}^y) - {\rm
      MEAN}(X_{{s_1},{s_2}}^y)}}{{{\rm STD}(X_{{s_1},{s_2}}^y)}},
\label{equ1}
\end{equation}
and,
\begin{equation}
W_{{s_1},{s_2}}^y = \frac{{{\rm MIN}(X_{{s_1},{s_2}}^y) - {\rm
      MEAN}(X_{{s_1},{s_2}}^y)}}{{{\rm STD}(X_{{s_1},{s_2}}^y)}}
\label{equ2}
\end{equation}
where ${\rm MAX}$ and ${\rm MIN}$ are the maximum and minimum values
of the cross-correlation function, ${\rm MEAN}$ and ${\rm STD}$ are
the mean and standard deviation, and the superscript $y$ denotes a
specific year.  Typical time series and their cross-correlation
functions are shown in Fig. \ref{fig1}. In this sample, the absolute
value of a minimal (negative) cross-correlation function is much
larger than the maximal value.

We implemented the above procedure for winter and summer time series,
separately. Specifically, we calculate the cross-correlation functions
of each pair of sites, where the data ranges from May 1st to Aug. 31st
(123 days) for southern hemisphere (SH) winter (northern hemisphere,
(NH) summer) or from Nov. 1st to Feb. 28th (120 days) for SH summer
(NH winter). We choose a maximal time lag of 72 days and a time period
of $y \in [1948,2010]$. The correlation coefficient and the link
weight are based on 12 months of data, in order to have sufficient
statistics. This is done by ``gluing'' together three consecutive
winters (summers), such that the total number of months is 12.

\begin{figure*}[htbp]
  \begin{minipage}[b]{0.68\columnwidth}
    \centering
    \includegraphics[width=\columnwidth]{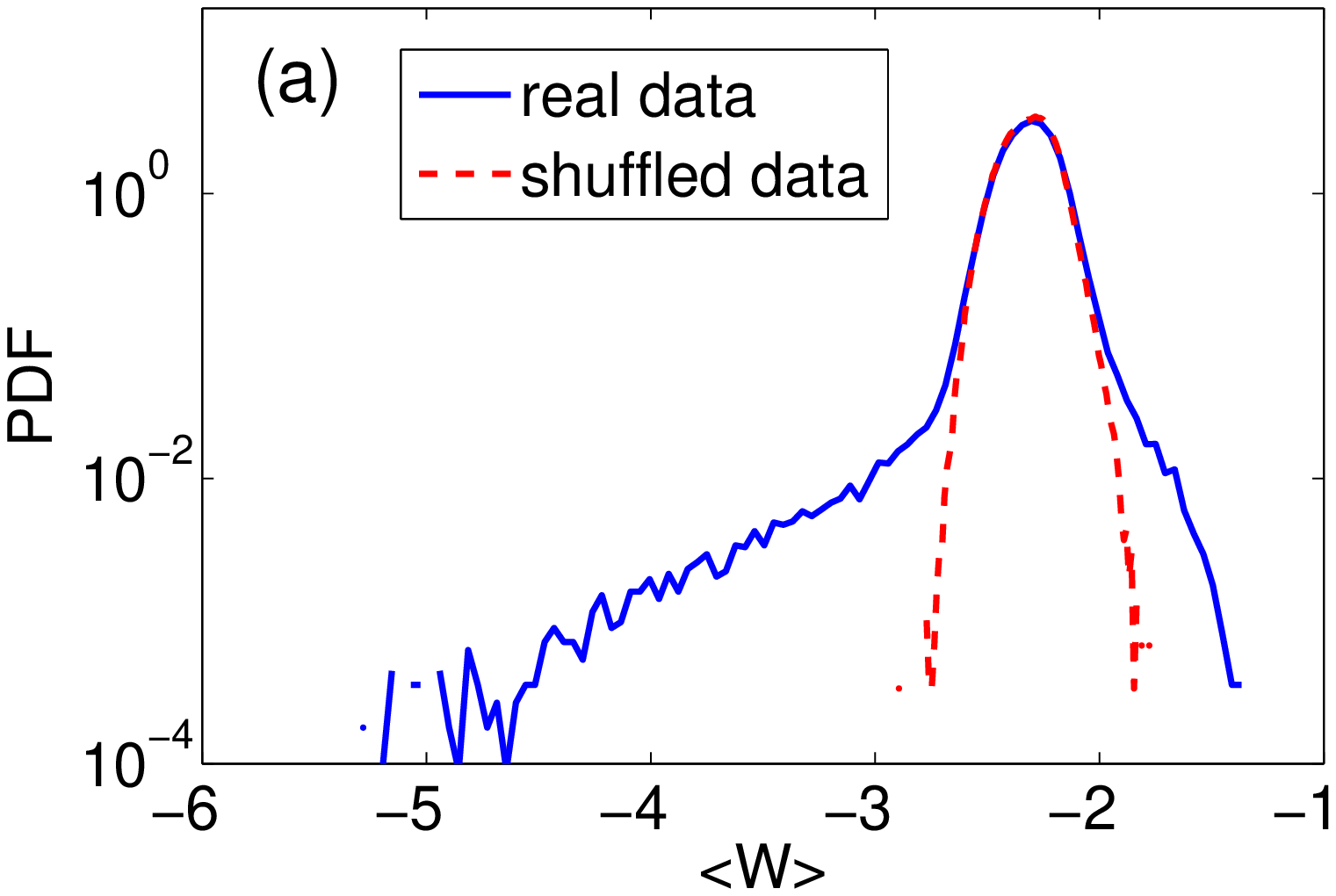}
  \end{minipage}%
    \begin{minipage}[b]{0.68\columnwidth}
    \centering
    \includegraphics[width=\columnwidth]{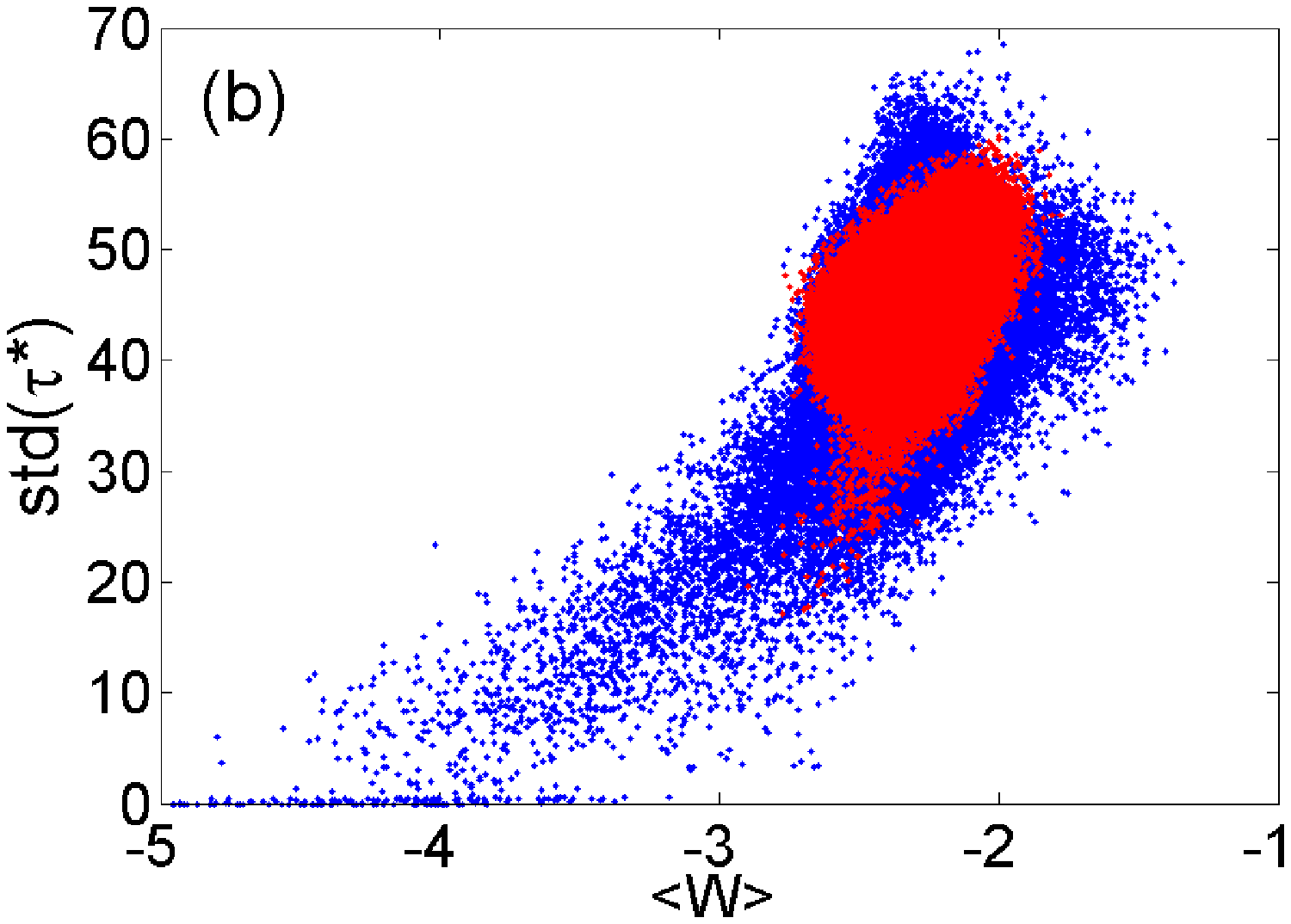}
  \end{minipage}%
    \begin{minipage}[b]{0.68\columnwidth}
    \centering
    \includegraphics[width=\columnwidth]{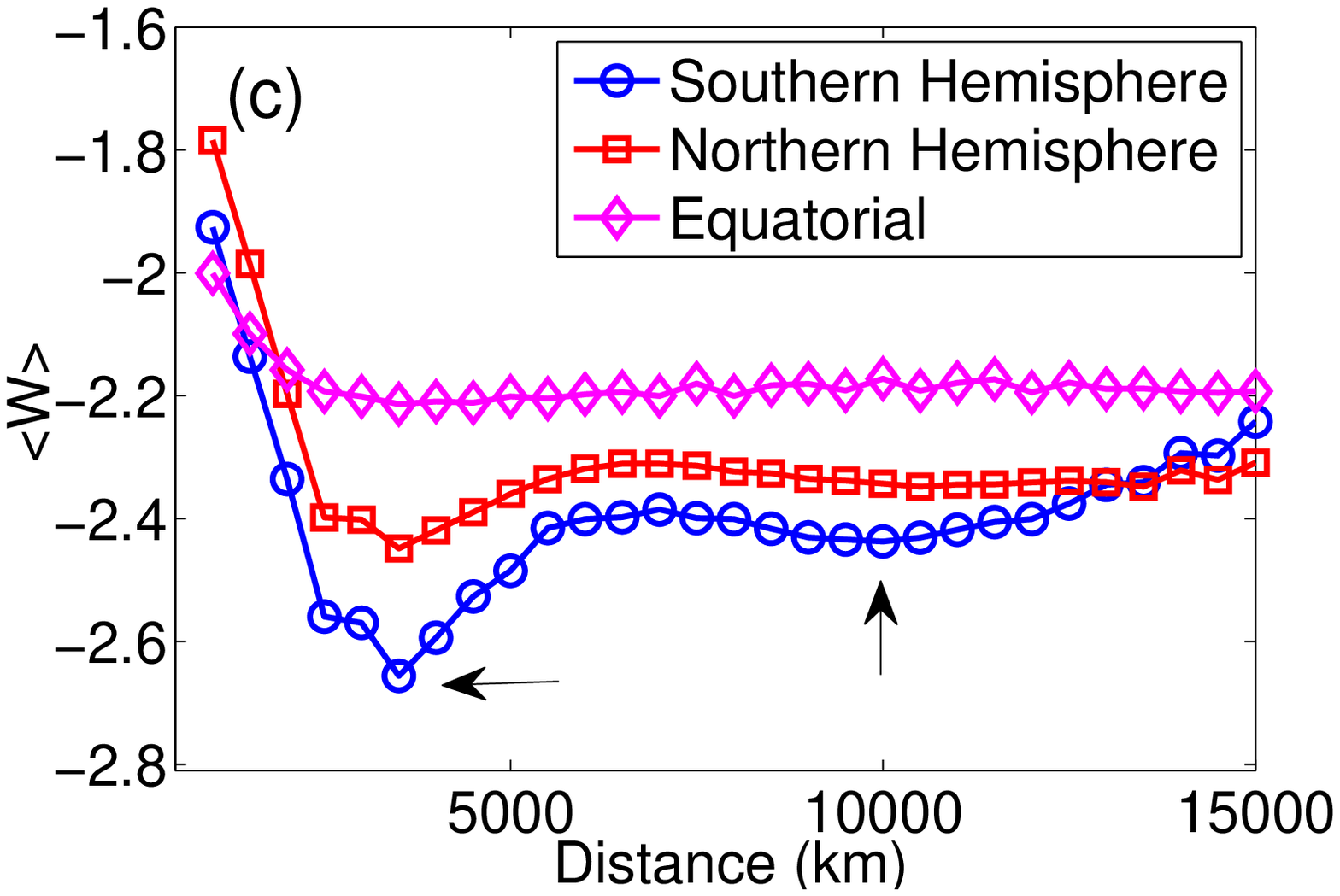}
  \end{minipage}
  \caption{(Color online) (a) The probability density function (PDF)
    of the mean weight of negative links in the globe, in both real
    (blue solid line) and shuffled (red dashed line) data. (b) The
    dependence of the standard deviation of time delay $\tau^*$ on the
    mean weight $\langle W \rangle$ for all possible links, for real
    (blue) and shuffled (red) data. (c) The mean negative link weight
    as a function of distance $d$, for the SH (blue circles), NH (red
    squares) and the equator (pink diamonds) regions. The figure is
    based on the globe temperature records at 1000 hPa isobar for
    Nov. to Feb..}
\end{figure*}
We analyze a near surface (1000 hPa) temperature time series. First,
we focus on the properties of link weight around the globe (Fig. 2)
for the months Nov. to Feb.. To identify the significant links we
apply a shuffling procedure in which the order of the years is
shuffled while the order within each year remains unchanged. This
shuffling scheme is aimed at preserving all the statistical quantities
of the data, such as the distribution of values, and their
autocorrelation properties, but omitting the physical dependence
between different nodes. Fig. 2(a) and (b) depict the link weight
statistics for the real and shuffled data. High negative mean link
weight values exist in the probability density function (PDF) of the
real data but are missing in the shuffled data, and therefore are not
likely to occur by chance. Moreover, high variability (std.) during
different years of the time delays of links $\tau^*$ is also a
signature of random behavior~\cite{Bashan2012} (Fig. 2(b)). The
differences between the distribution of real data and shuffled data
indicate that many significant negative links exist in the climate
network (Fig. 2(a),(b)). We obtain similar results for positive links
and other fields (not shown).

Next, we divide the world into three geographical zones, the SH (from
22.5\degree{S} to the south pole), the NH (from 22.5\degree{N} to the
north pole) and the equator (between 22.5\degree{S} and
22.5\degree{N}). We then calculate the (geographical and temporal)
mean link weight $\langle W_{s_1,s_2}\rangle$ as a function of the
geographical distance of links, $d$. It is clear that in the SH, there
is a preferred distance of $\sim$ 3,500 km and a much weaker one of
$\sim$ 10,000 km (Fig. 2(c)). In the NH region, we find a similar,
weaker dependence, while in the equatorial region, there is no
preferred distance (Fig. 2(c)). These preferred distances may be
associated with atmospheric Rossby
waves~\cite{Chang2005,Chang1999(1),Chang1999,Chang1993}, which have a
wavelength of $\sim$ 7,000 km and which are known to be pronounced in
the SH, weaker in the NH and absent in the equatorial
region~\cite{Chang1999(1),Chang1999}. The negative peaks at 3,500~km
and 10,000~km represent a 1/2 wavelength and a 3/2 wavelength of the
observed Rossby wavelength.

\begin{figure}[htb]
\center\includegraphics[width=0.55\textwidth]{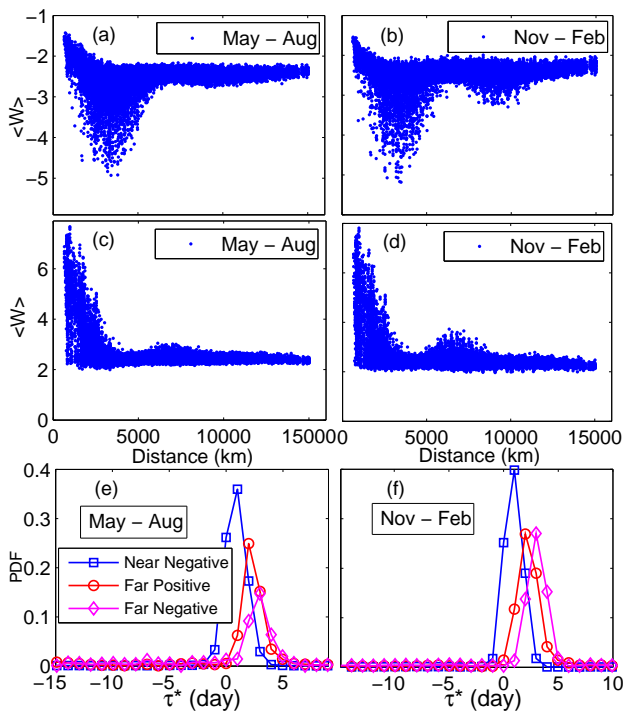}
  \caption{(Color online) The dependence of the weight of negative
    links on the distance $d$ in the SH during (a) winter and
    (b) summer months. (c),(d) Same as (a),(b) for positive links. The
    PDF of time delays $\tau^*$ of near negative (blue squares,
    $d\sim$ 3,500 km), far positive (red circles, $d\sim$ 7,000 km),
    and far negative (pink diamonds, $d\sim$ 10,000 km) weighted
    links, in the SH during (e) winter and (f) summer. The figure
    is based on the temperature field at 1000 hPa isobar.}
\end{figure}

To further consolidate the association of the observed pattern in the
climate network with Rossby waves, we compare the seasonality of this
pattern with the known seasonal characteristics of Rossby waves. In
Fig. 3(a)-(d), we plot the negative and positive weights of all SH
links, for the winter and summer months separately. Each point
represents an average link weight $\langle W \rangle$ over years
versus its distance $d$. The negative weights, defined in
Eq.~(\ref{equ2}), have a pronounced enhanced distribution of large
weights for $d \sim$ 3,500 km during both summer and winter months,
while for the SH summer months (Nov. to Feb.), there is an additional
preferred distance of $\sim$ 10,000 km (Fig. 3 (b)), both in
accordance with the 1/2 and 3/2 wavelengths of atmospheric Rossby
waves. Around the full wavelength distance (i.e., $d \sim 7,000$~km),
we find, as expected, an enhanced distribution of large positive
weights (defined in Eq.~(\ref{equ1})) (Fig. 3 (c),(d)). However, a
larger abundance of strong waves (represented by links at 1/2, 1, and
3/2 wavelengths) during summer, in comparison to winter months, is
clearly seen in Fig. 3 (a)-(d), in agreement with the clearer pattern
of Rossby waves found during the SH summer in the
SH~\cite{Chang1999(1),Chang1999,Chang1993}. The situation is similar
in the NH (not shown), for the NH winter. In the distribution of
positive weights (Fig. 3(c),(d)), one clearly sees the links that
emerge around $d < 2,000$~km due to the proximity effect mentioned in
the introduction.


Atmospheric Rossby waves have a characteristic group velocity, and we
now estimate it based on the climate network results shown above. To
achieve this, we divide $d$ by $\tau^{*}$ for each link, under the
assumption that $\tau^{*}$ is a good estimate for the underlying
dynamical delay between the two sites (nodes) \cite{cimp}. Since
$\tau^{*}$ is only meaningful for links with weights above the
background noise level (see Fig. 2(b), and ref.~\cite{Berezin2012}),
we limit the analysis to links with weights $|\overline{W}| > 2.8$
~\cite{fn1}. Furthermore, to avoid links that are prone to the
proximity effect (as such links cannot easily fit in the current
Rossby wave interpretation), when considering positively correlated
links, we consider links with $d > 5,000$ km. Also, we constrain the
near negative links to be within $d \in [2,000 km, 5,000 km]$ and far
negative links to have $d > 8,000$ km. The PDFs of time delay
$\tau^{*}$ of the near negative links, the far positive links and the
far negative links are shown in Fig. 3(e),(f). By our convention,
positive $\tau^{*}$ means an eastward energy flow, which is the
typical case for most observed links. The results point to a time lag
of $\tau\approx$ 1 day for the near negative links (i.e., links in the
first peak in Fig. 3(a)), $\tau \approx$ 2-3 days for the far (beyond
the proximity effect) positive links, and $\tau\approx$ 3-4 days for
the far negative links (i.e., belonging to the second peak in the
scatter plot of Fig. 3(a,b)). Based on these numbers, our estimated
group velocities are in the range $[20-35]$m/s, consistent with the
range $[23-32]$m/s reported in previous
studies~\cite{Chang1999,Berbery1996}.

\begin{figure*}[htbp]
\center\includegraphics[width=0.75\textwidth]{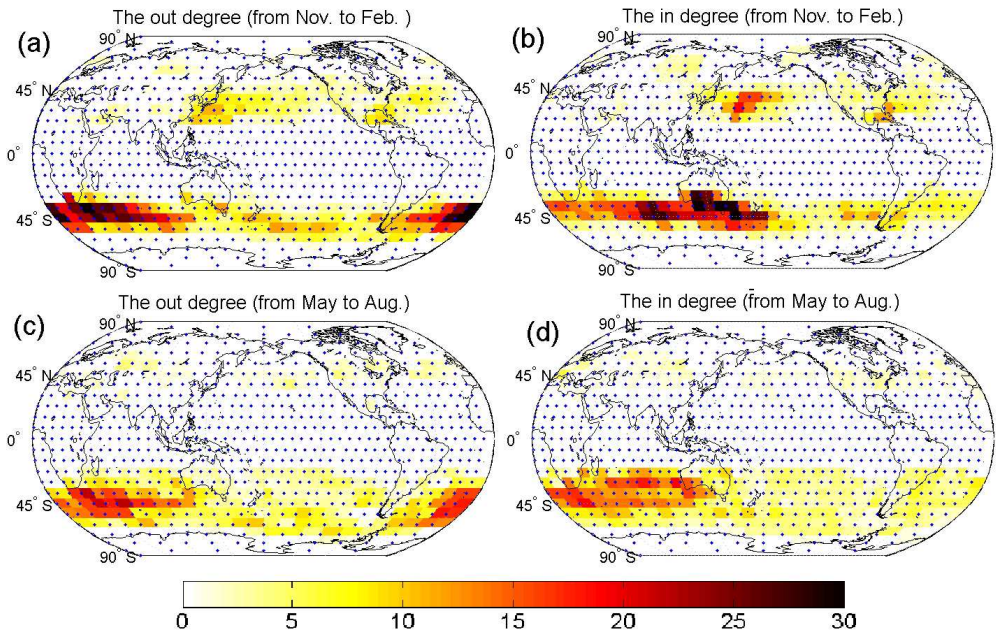}
  \caption{(Color online) The out- (a) and in-degree (b) of the
    climate network structure from Nov. to Feb. (SH summer). The
    out- (c) and in-degree (d) of the climate network structure from
    May to Aug. (SH winter). The figure is based on the temperature
    record at 1000 hPa isobar.}
\end{figure*}

The climate network has a unique geographical structure that can be
compared with the geographical structure of Rossby waves. Since the
network is directed---positive $\tau$ indicates eastward flow while
negative $\tau$ indicates westward flow---we distinguish between a
link that is pointing toward a node (where the number of links
pointing to a specific node is referred to below as ``in-degree''), or
away from the node (referred to below as ``out-degree'') (see
~\cite{Gozolchiani2011}). Fig. 4 depicts the mean in- and out-degrees
of each node, excluding the equatorial region that conforms with a
pattern that is not related to our current discussion. The observed
structure is consistent with the structure of Rossby
waves~\cite{Chang1999}. First, the wave band in the SH from May to
August (SH winter, Fig. 4(c),(d)) is broader than that of the SH
summer (Nov. to Feb., Fig. 4 (a,b)). Second, the atmospheric Rossby
wave structure in the NH summer is less pronounced. Third, the wave
structure in the SH summer (Nov. to Feb.) lies on a band centered near
50\degree{S}. All the above characteristics are consistent with the
properties of Rossby waves.

\begin{figure}[htb]
\center\includegraphics[width=0.5\textwidth]{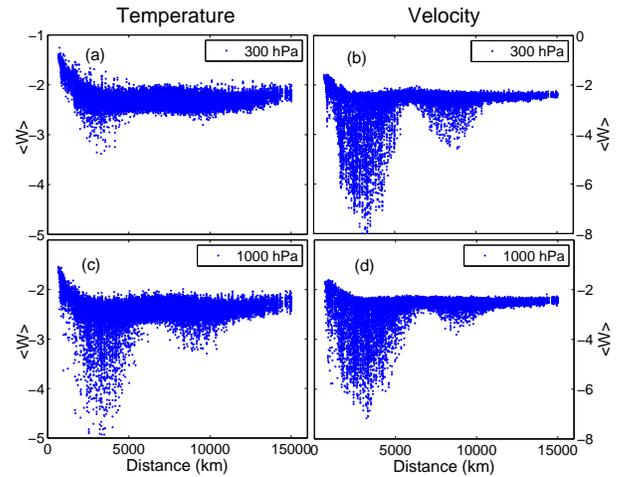}
  \caption{(Color online) The dependence of the weight of negative
    links on the distance $d$ in the SH during SH summer months for
    temperature (left panels) and meridional velocity (right panels)
    for 300 hPa (upper panels) and 1,000 hPa (lower panels).}
\end{figure}

Previous studies found the 300 hPa meridional velocity field to be the
most suitable for studying the characteristics of Rossby waves
\cite{Chang1993,Chang1999,Chang1999(1)}. Our method captures the wave
properties also using other fields at various altitudes. In
particular, we showed above that the wave pattern is clearly seen at a
ground level (1000 hPa) temperature field, a more common and reliable
variable.  In Fig. 5, we compare, using the climate network technique,
between the mean weight distributions of the negative links of the
meridional velocity and temperature fields. We find that the pattern
of the two fields is similar, although the meridional velocity yields
larger weights. In addition, the meridional velocity yields a clearer
pattern at the high altitude of 300 hPa, while the temperature field
yields a clearer pattern at the ground level of 1000 hPa.


In summary, we analyze the properties of the climate network by
considering, separately, positive and negative correlations
(links). The most dominant links in the climate network with a
geographical distance larger than 2,000 km are found for distances of
$\sim$ 3,500 km, $\sim$ 7,000 km and $\sim$ 10,000 km. These distances
coincide with the 1/2, 1 and 3/2 wavelengths of common atmospheric
Rossby waves. Moreover, the time delays associated with these
distances are in agreement with the direction of the energy flow and
with the group velocity of the atmospheric Rossby waves. The
pronounced length scales of the climate network, the dominance in the
SH in comparison with the NH, and the dominance during the SH summer
in the SH are all consistent with the properties of atmospheric Rossby
waves. All of these factors thus provide strong support for the
association of the majority of the climate network far links with
Rossby waves.

\section*{ACKNOWLEDGMENTS}

The authors would like to acknowledge the support of the LINC project
(no. 289447) funded by the EC's Marie-Curie ITN program
(FP7-PEOPLE-2011-ITN) and the
Israel Science Foundation for financial support. We thank Edmund Chang and Nili Harnik for their
assistance in understanding the characteristics of Rossby waves.


\begin{thebibliography}{99}
\bibitem{comment1}{The original resolution of the reanalysis data is
    2.5\degree{}. The number of nodes that were analyzed was
    normalized by the cosine of latitude, to keep the density of nodes
    equally distributed.}
\bibitem{fn1}{We have checked thresholds in the range $2.7 - 3$
    without significant qualitative changes in the results.}
\end{thebibliography}
\end{document}